\documentclass[final,5p,times,twocolumn]{elsarticle}

\usepackage{epsfig}
\usepackage{graphicx}
\usepackage{amssymb}
\usepackage{comment}

 \biboptions{comma,sort&compress}

\journal{Physics Letters B}
\usepackage{ulem}
\def\nuc#1#2{\relax\ifmmode{}^{#1}{\protect\text{#2}}\else${}^{#1}$#2\fi}
\newcommand{\hepb}{\nuc{6}{He}+\nuc{208}{Pb} }

\begin{document}

\begin{frontmatter}



\title{Simultaneous analysis of elastic scattering and transfer/breakup channels for the \nuc{6}{He}+\nuc{208}{Pb} reaction  at energies near the Coulomb barrier}  


\author[famn,cna]{J. P. Fern\'andez-Garc\'ia} 
\ead{jpfernandez@us.es}
\author[famn,cna]{M. A. G. Alvarez} 
\ead{malvarez@us.es}
\author[famn]{A. M. Moro}
\ead{moro@us.es}
\author[famn,csic]{M. Rodr\'{\i}guez-Gallardo}
\ead{mrodri@us.es}

\address[famn]{Departamento de FAMN, Universidad de Sevilla, Apartado 1065, 41080 Sevilla, Spain.}
\address[cna]{Centro Nacional de Aceleradores, Universidad de Sevilla, 41092 Sevilla, Spain.}
\address[csic]{Instituto de Estructura de la Materia, CSIC, Serrano 123, 28006 Madrid, Spain.}

\begin{abstract}
The elastic and $\alpha$-production channels for the  \nuc{6}{He}+\nuc{208}{Pb} reaction are investigated at energies around the Coulomb barrier 
($E_\mathrm{lab}$=14, 16, 18, 22, and 27 MeV). 
The effect  of the two-neutron transfer channels on the elastic scattering has been studied within the Coupled-Reaction-Channels (CRC) method.  We find that the explicit inclusion of these channels allows a simultaneous description of the elastic data and the inclusive $\alpha$ cross sections  at backward angles. Three-body 
Continuum-Discretized Coupled-Channels (CDCC) calculations are found to reproduce the elastic data, but not the transfer/breakup data. The trivially-equivalent local polarization potential (TELP) derived from the CRC  and CDCC calculations are found to explain the features found in 
previous phenomenological optical model calculations for this system. 
\end{abstract}


\date{\today}

\end{frontmatter}


\section{Introduction}
\label{sec:intro}

Since the early nineties, a considerable amount of experimental data of reactions induced by the Borromean
nucleus \nuc{6}{He} on a variety of targets has been accumulated. In the energy regime around the Coulomb barrier, and 
for medium-mass and heavy targets, these data 
show some common remarkable features. First, the elastic scattering angular distribution  does not follow the 
expected Fresnel pattern, that characterizes the scattering of heavy ions at these energies. Instead, the 
characteristic Fresnel peak is partially or completely 
suppressed and the angular distribution shows a smooth decrease as a function of the scattering angle.
Second,  these reactions exhibit a large yield of $\alpha$ particles \cite{Agu00,DeY05,DiPiet04,Nav04,Raa04}. This is clearly a consequence of the weak 
binding of the \nuc{6}{He} nucleus, which can be easily broken up by the strong couplings induced by the target. 

Heavy ion reactions between stable nuclei can be understood in terms of the {\it strong absorption} picture \cite{Sat83}. This implies the existence of an interaction 
distance ($R_{sa}$) such that impact parameters  below $R_{sa}$ are dominated by strong absorbing processes (i.e.\, deep inelastic 
collisions, compound nucleus, etc),  whereas for separations 
larger than  $R_{sa}$, the nuclei barely interact, and elastic scattering dominates. Direct reactions, such as nucleon transfer or diffractive breakup,  take place  only for peripheral or grazing collisions.
 However, the analysis of recent experiments suggest that this picture may be inadequate to interpret collisions  involving weakly bound nuclei. In the \hepb case, optical model (OM) calculations performed at Coulomb-barrier energies \cite{Kak03,Kak06,San08} show that, in order to reproduce 
the elastic data, one requires  optical potentials with a very diffuse imaginary tail. If the interacting  potential 
is parametrized using a standard complex Woods-Saxon form, the diffuseness parameter of the imaginary component required to reproduce the elastic data  turns out to be around $a_i \approx 2$~fm, a value that is considerably larger than the diffuseness derived  
from the matter distribution ($\approx 0.6$~fm).  This phenomenon, recently referred to as \textit{long range absorption} effect, suggests the presence of reaction channels that remove flux from the elastic channel at distances 
well beyond the strong absorption radius, in contrast to the picture suggested by the strong absorption model.  Continuum-Discretized Coupled-Channels (CDCC) calculations using either a simplified \textit{di-neutron} model  \cite{Rus05,Mor07} or a more realistic three-body model for the 
 \nuc{6}{He} nucleus \cite{Mat04,Mat06,Cres07a,Cres07b,Rod07,Rod08,Rod09} indicate 
that this long range absorption phenomenon can be explained in terms of the strong couplings to the breakup channels due mainly to the dipole Coulomb interaction. 

Despite their success to describe the elastic data angular distributions, it was shown in \cite{Esc07} that these  CDCC calculations failed 
to reproduce the energy and angular distributions of the $\alpha$ particles observed at large angles in the same experiment. On the other hand, 
it was shown in the same work that these alpha particles distributions could be well accounted for by assuming a two-neutron 
transfer mechanism leading to very excited states of the target. These calculations were performed using the  
distorted-wave Born approximation (DWBA), assuming a two-neutron mechanism leading to both bound and unbound final states of the target.  The relevance of the one- and two-neutron transfer channels has been evidenced in other reactions involving   \nuc{6}{He} \cite{Agu01,Raa04,Nav04,DiPiet04,DeY05}.

The main difference between the CDCC and DWBA calculations is due to the couplings included in each method. This 
is schematically illustrated  in Fig.~\ref{fig:couplings}. The CDCC method assumes a  {\it direct breakup} mechanism in which the breakup process is treated as an inelastic excitation of the projectile to its continuum spectrum (top panel). On the other hand, the
DWBA method (or more elaborated forms of the transfer amplitude \cite{Mor06a}) is based on a {\it transfer to the 
continuum} picture, which treats the \nuc{6}{He} breakup assuming a neutron 
transfer mechanism populating  \nuc{210}{Pb}* states, as shown in the bottom panel of Fig.~\ref{fig:couplings}. Since both the projectile 
and target states form a complete set, we expect both methods to give similar results provided that the underlying interactions are the same and 
the model space is sufficiently large to achieve convergence of the studied observables \cite{Mor06a}. In this respect, the  direct breakup and transfer 
to the continuum methods can be regarded as alternative  methods to describe the projectile dissociation.
 Therefore, the failure of the CDCC approach to describe the 
$\alpha$ yield could be attributed to the truncation of the model space required in practical calculations, 
and/or to the neglect of the transfer to the bound states of the target which, at these energies, may contribute 
significantly to the $\alpha$ cross section. 

The transfer to the continuum calculations presented in \cite{Esc07} are based on the DWBA approximation and describe 
the elastic channel by means of a phenomenological OM, with parameters adjusted to the elastic data. The features of this 
OM potential must reflect the effect of channel couplings on the elastic channel.  Since the two-neutron transfer channels were found 
to be very important to describe the $\alpha$ yields, it would be desirable to investigate to what extent  the 
explicit inclusion of these transfer couplings can explain simultaneously the observed features of the elastic scattering and the 
$\alpha$ channel, without the need of a phenomenological OMP adjusted to the data. To study the effect of the transfer channels on the elastic channel one needs to go beyond the Born approximation, that is, 
to use the  Coupled-Reaction-Channels (CRC)  method. 

With this purpose in this work we present CRC and CDCC calculations for the \nuc{6}{He}+\nuc{208}{Pb} reaction at near-barrier energies, 
comparing the calculated elastic and $\alpha$ cross sections with the available data. 
In the CRC calculations we  rely on the scheme illustrated in the bottom panel 
of  Fig.~\ref{fig:couplings}, but we allow  back couplings from the transfer to the elastic channels. We show 
that, unlike the CDCC, the CRC formalism is able to explain simultaneously  the long-range absorption effect on the elastic scattering and the large yield of $\alpha$ particles for this reaction. 
 
Although the  calculations are restricted to the  \nuc{6}{He}+\nuc{208}{Pb} case, we believe that the conclusions can be extrapolated to other  reactions induced by weakly bound projectiles. 

\section{\label{sec:cdcc} CDCC calculations}

\begin{figure}[h]
 \vspace{0.3cm}
  \hfill
  \begin{minipage}[t]{.45\textwidth}
    \begin{center}  
      \epsfig{file=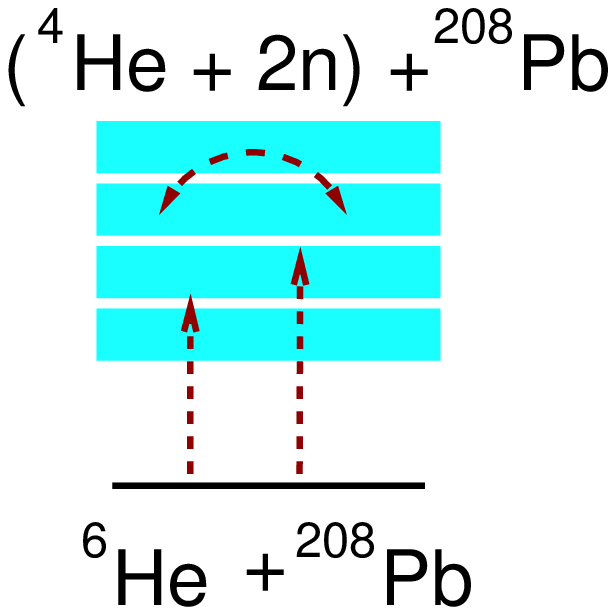, scale=0.5}
    \end{center}
  \end{minipage}
  \hfill
  \begin{minipage}[t]{.5\textwidth}
    \begin{center}
      
       \epsfig{file=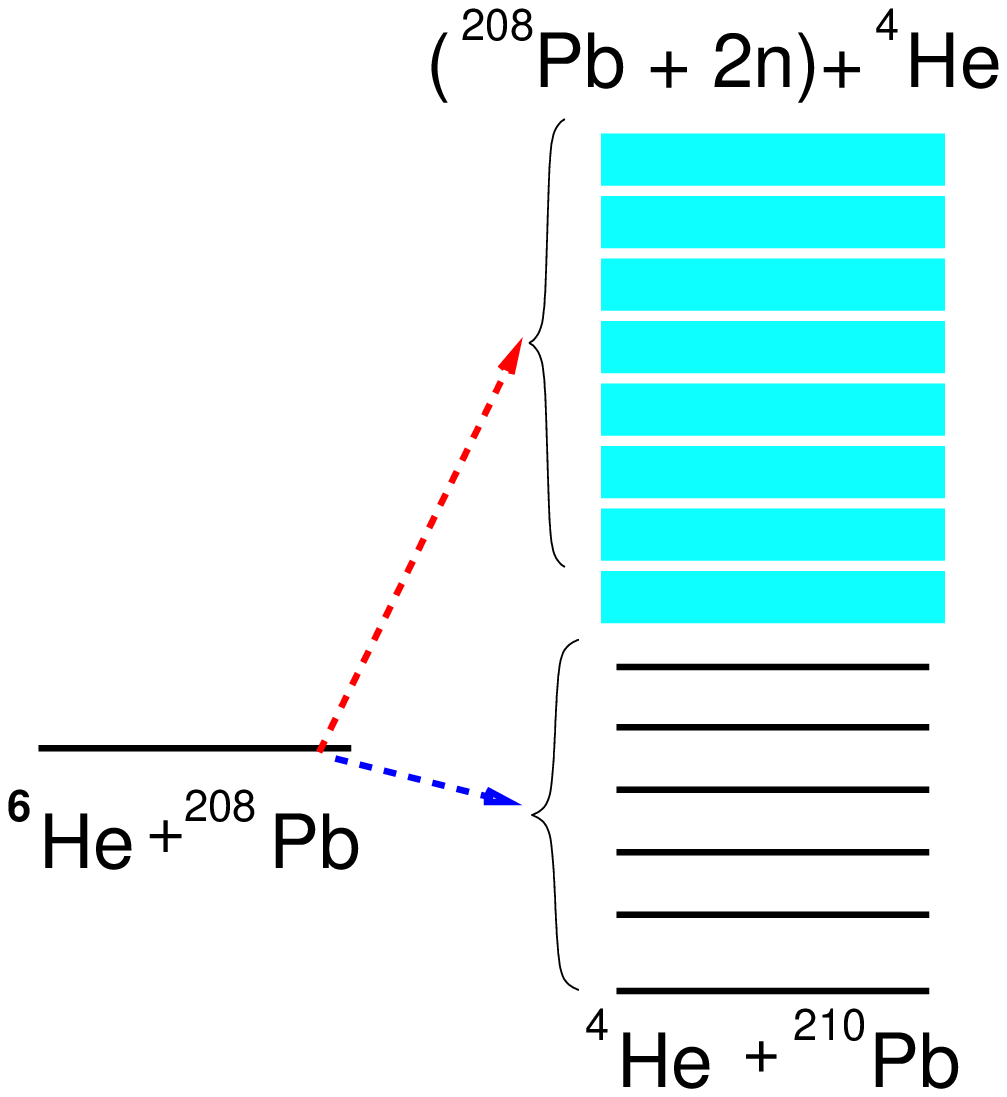, scale=0.5}
    \end{center}
  \end{minipage}
  \caption{\label{fig:couplings}  Schematic representation of the couplings included in the direct breakup (CDCC)
             (top) and  transfer to the continuum (DWBA, CRC) (bottom) calculations.}
  \hfill
\end{figure}

Within the Continuum-Discretized Coupled-Channels (CDCC) method, the dissociation of the projectile is  
treated assuming a direct breakup picture, in which the  projectile is excited to its unbound states  (see Fig.~\ref{fig:couplings}). For \nuc{6}{He} reactions, 
this has been done using a three-body reaction model (based on a simple di-neutron  model of the projectile, \nuc{4}{He}+2n  \cite{Rus05,Mor07}) or 
a four-body reaction model (using a realistic  three-body model of \nuc{6}{He}=\nuc{4}{He}+n+n \cite{Rod07,Rod08,Rod09,Mat04,Mat06}). In this work, we rely on the first model because this will permit a more meaningful comparison with the CRC 
calculations presented in Section \ref{sec:crc}, which are also based on a two-body model of the \nuc{6}{He} nucleus. In particular,  we will make 
use of the improved di-neutron model proposed in \cite{Mor07,Esc07}, in which the \nuc{4}{He}+2n relative wavefunctions are calculated in a Woods-Saxon 
potential with radius $R=1.9$~fm and diffuseness parameter $a_0=0.39$~fm. The ground state wavefunction is calculated assuming a $2S$ 
configuration and an effective  separation energy of the two-neutron cluster  $S_{2n}=1.6$~MeV. For the $\alpha$+\nuc{208}{Pb}  
interaction we took the potential of Barnett and Lilley \cite{Bar74}, whereas the 2n+\nuc{208}{Pb} interaction was approximated by the 
deuteron-\nuc{208}{Pb} global potential of  Ref.~\cite{DCV80}.   \nuc{6}{He} continuum states with relative angular momentum  $l_i=0,1,2$ 
 for the  \nuc{4}{He}+2n relative motion were considered. For the $l_i=2$ continuum, the potential depth was adjusted to 
reproduce the known $2^+$ resonance at $E_x=1.8$~MeV above the g.s.  For each value of $l_i$, the continuum was discretized using the standard 
binning method. The maximum excitation energy depended somewhat on the incident energy, ranging from 5 MeV (for $E_\mathrm{lab}$=14 MeV) 
to 8 MeV (for $E_\mathrm{lab}$=27 MeV). These calculations were performed with the code {\sc fresco} \cite{Tho88}.

The calculated elastic differential cross sections are compared  in Fig.~\ref{he6pb_el_crc} (dashed lines) with the experimental 
data from Refs.~\cite{Kak03,Kak06,San08}.  Despite the simplified structure model used for the \nuc{6}{He} nucleus and the absence of any 
free adjustable parameter, the overall agreement with the data is 
very good at the five  considered energies.   
 Below the barrier ($E_\mathrm{lab}$=14, 16, and 18 MeV), 
the inclusion of the breakup couplings produces a decrease of the cross section  with respect to Rutherford at c.m.\ angles beyond 60$^\circ$. Above the 
barrier ($E_\mathrm{lab}$=22 and 27 MeV) the main effect is the absence 
of the Fresnel peak. 

In Figs.~\ref{dsde_crc} and \ref{dsdw_crc}, the calculated energy and angular distributions of the $\alpha$ particles (dotted lines) are 
compared with the data from Ref.~\cite{Esc07} (solid circles). Clearly, the calculations fail to describe both observables and, therefore, 
we conclude that the direct breakup model, at least within this restricted model space, cannot explain the large yield of $\alpha$ particles produced at large angles.  On the other hand,  it was shown in Ref.~\cite{Esc07} that these distributions
could be well reproduced by means of DWBA calculations, assuming a two-neutron transfer mechanism.


\section{\label{sec:crc} CRC calculations}
In the distorted-wave Born approximation (DWBA) calculations of  Ref.~\cite{Esc07}, the entrance channel was described using a phenomenological OM potential, with parameters adjusted to reproduce the experimental elastic angular distribution.
In the Coupled-Reaction-Channels (CRC) calculations,  we aim at  explaining the features of the elastic scattering as a consequence of the same couplings which are responsible for the production 
of the  $\alpha$ particles with the objective of reproducing both channels within the same framework. Therefore, 
in order to avoid double counting, in the CRC calculations  one has to use instead a {\it bare} 
interaction, that is, a  potential that represents the interaction between the colliding nuclei in absence 
of couplings that will be explicitly included.  

\begin{figure}[tb]
\begin{center}
 {\centering \resizebox*{0.95\columnwidth}{!}{\includegraphics{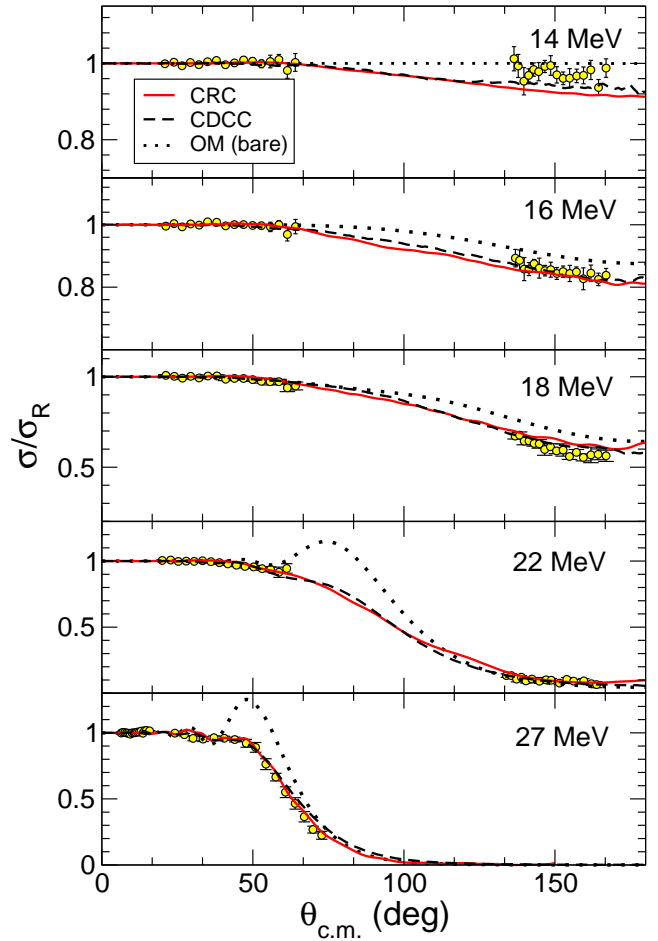}} \par}
\caption{\label{he6pb_el_crc} Elastic scattering angular distribution as a function of the c.m.\ scattering angle  
for the reaction $^{6}$He+$^{208}$Pb at $E_{\rm lab}$=14, 16, 18, 22, and 27 MeV. The dotted line is the one-channel calculation 
with the bare interaction, $U_\mathrm{SPP}(r)$, the dashed line is the three-body CDCC calculation and the  solid line is the full CRC calculation.  The  circles are the experimental data from Ref.~\cite{Kak03,San08}.}
\end{center}
\end{figure}

In particular, we use for the bare interaction  the S\~ao Paulo potential (SPP) \cite{Cha02,Cha97}. This is a
microscopic potential obtained by means of a double-folding procedure, using the matter densities of the colliding nuclei and an 
effective  nucleon-nucleon interaction, multiplied by an energy-dependent term, which  accounts for part of the non-locality of the 
optical potential. This provides a parameter-free prescription, which has been successfully applied to a large variety of systems \cite{Cha97,Cha98,Gal98,Cha02,Alv03,Alv99,Sil01,Ros02}. Indeed, the choice of the bare interaction is not unique. The present choice is  convenient because, being based on the matter densities of the colliding nuclei, reduces many ambiguities associated with more phenomenological potentials.  The bare interaction includes also a short-range  imaginary potential of Woods-Saxon shape and parameters  $W=50$~MeV, $R_0=1$~fm, $a_0$=0.1~fm, to simulate the in-going boundary condition for fusion.  This bare interaction was recently used in the OM analysis performed in \cite{jp10} for the same elastic data.  As it was shown in that work, the bare interaction alone is unable to reproduce the elastic 
data. This is illustrated in Fig.~\ref{he6pb_el_crc}, where 
the one-channel calculation with the bare interaction (dotted line) is compared with the experimental data.  
At $E_\mathrm{lab}=22$ and 27 MeV (above the nominal Coulomb barrier) this calculation predicts a pronounced rainbow, which is absent from the data. Below the barrier ($E_\mathrm{lab}=14$, 16, and 18~MeV), the calculation overestimates the data for angles beyond 50$^\circ$. Thus, 
this prescription fails to describe the elastic data both below and above the Coulomb barrier. The failure is clearly due to the influence of channel couplings, mainly, the 2n removal channels, which are expected to 
be very important for a loosely bound nucleus like \nuc{6}{He}. In Ref.~\cite{jp10}, the effect of these channels  on the elastic channel was taken into account 
adding a phenomenological component to the bare interaction of Woods-Saxon form. It was found that the real and imaginary parts of the Woods-Saxon potentials required a large diffuseness  parameter in order to reproduce the data.



 In the present work, 
we start from the same bare interaction but, instead of adding any phenomenological component, we introduce the 
two-neutron transfer couplings explicitly, using the CRC formalism.
 Using the 
prior representation, the transfer couplings involve a matrix element of the  operator $V_\mathrm{[2n+Pb]}+U_\mathrm{[\alpha+Pb]}-U_\mathrm{bare}$, where $V_\mathrm{[2n+Pb]}$ is a real interaction describing two neutron states of the 2n+\nuc{208}{Pb} system, $U_\mathrm{[\alpha+Pb]}$ is the optical 
potential describing the $\alpha$+\nuc{208}{Pb} elastic scattering and $U_\mathrm{bare}$  is the bare potential defined above,
 given by the S\~ao Paulo prescription.  In the CRC calculations we took the same  $\alpha$+\nuc{208}{Pb} and   2n+\nuc{208}{Pb} 
interactions used in the CDCC calculations of the preceding subsection,  in order to have a
meaningful comparison between both formalisms. Nevertheless, for the latter only the real part of the interaction 
was considered, in order to allow the inclusion of two-neutron bound states in \nuc{210}{Pb}.  
For the \nuc{6}{He} nucleus, we adopt the same
 di-neutron model used in the CDCC calculations. 
 For the  \nuc{210}{Pb}* final states, we considered several values  for the angular momentum of the relative angular motion for  
\nuc{208}{Pb}+2n ($l_f$). For each value of $l_f$, the energy spectrum was described by a set of representative states, following the procedure described in \cite{Esc07}. These representative states include both bound and unbound states, with 
respect to the two-neutron breakup threshold. Above this threshold, the 2n-\nuc{208}{Pb} continuum was discretized using 1 MeV bins, up to a maximum relative energy of 5 MeV. For energies below the threshold, we considered  5 bound states, evenly spaced in steps of 2 MeV.   The number of partial waves $l_f$ was progressively increased until 
convergence of the observables was found, in the kinematic region where comparison with data is possible (see discussion below).
The coupled equations were solved iteratively using the code {\sc fresco} \cite{Tho88}, until the desired degree 
of convergence of the elastic and transfer cross sections was achieved.  We include also the  non-orthogonality correction \cite{Sat83}. 

The elastic differential cross sections obtained with these CRC calculations are displayed in Fig.~\ref{he6pb_el_crc} with solid lines. For all the considered energies, the agreement with the CDCC calculation and hence with the data is significantly improved with respect to the single-channel calculation performed with the bare interaction alone. Therefore, the inclusion of the transfer channels produces essentially the same effect on the elastic cross section as the inclusion of the $^6$He continuum states in the CDCC calculations.   This is an important result which supports the 
conclusion that both, CDCC and CRC methods, populate to a large extent the same final states, although these states are  expressed in different  basis representations.   

Despite the good agreement between both methods, the rate of convergence is very different. The CRC calculations were performed with $l_f=0-8$ partial waves for the 
2n-\nuc{208}{Pb} relative motion, whereas the three-body CDCC calculations required only $l_i=0,1,2$ partial waves for 
2n-\nuc{4}{He}  in order to achieve convergence of the elastic scattering \cite{Mor07}. This indicates 
that the elastic scattering is mostly affected by the coupling to continuum states with small energy and angular momentum between $2n$ and $\alpha$, 
and hence a  representation based on the continuum of the projectile is more efficient to describe this 
observable.

\begin{figure}[tb]
\begin{center}
 {\centering \resizebox*{0.9\columnwidth}{!}
 {\includegraphics{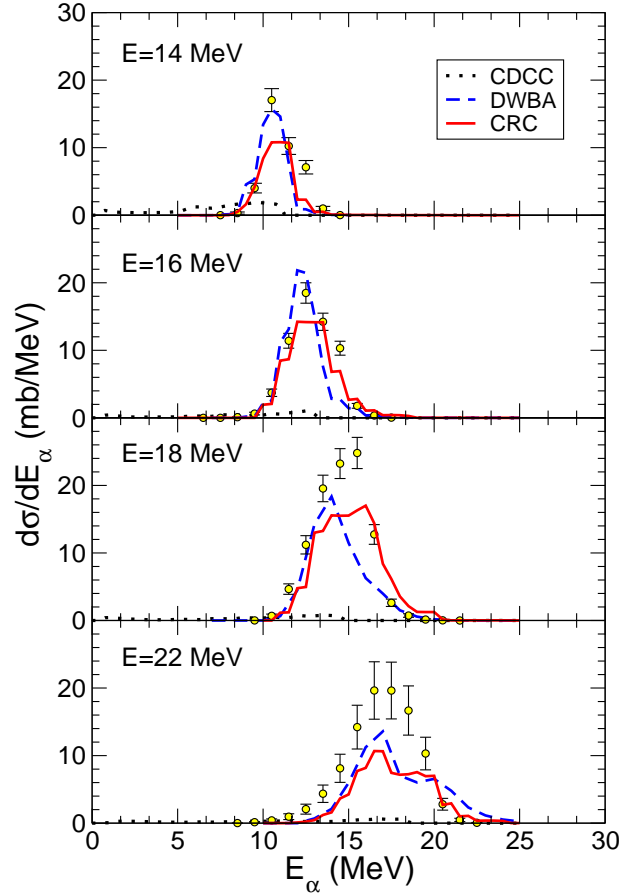}} \par}
\caption{\label{dsde_crc} Energy distribution of the $\alpha$ particles produced in the reaction $^{6}$He+$^{208}$Pb at $E_{\rm lab}$=14, 16, 18, and  22 MeV, integrated in the angular interval $\theta_\mathrm{lab}=132^\circ-164^\circ$. The dotted and solid lines are the CDCC and CRC calculations. The dashed lines and circles are, respectively, the DWBA calculations and the experimental data from Ref.~\cite{Esc07}.}
\end{center}
\end{figure}

\begin{figure}[tb]
\begin{center}
 {\centering \resizebox*{0.9\columnwidth}{!}
 {\includegraphics{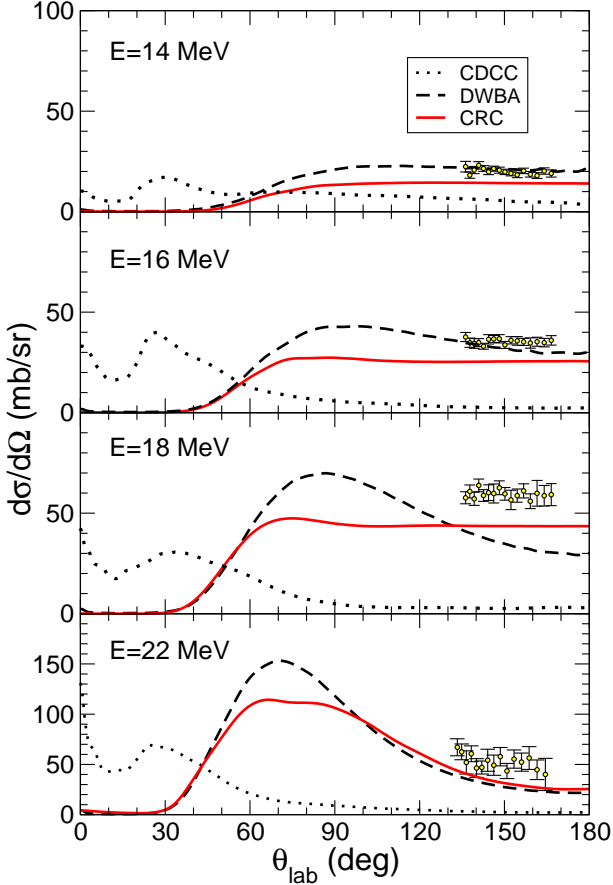}} \par}
\caption{\label{dsdw_crc} Angular distribution of the $\alpha$ particles produced in the reaction $^{6}$He+$^{208}$Pb at $E_{\rm lab}$=14, 16, 18, and 22 MeV. The dotted and solid lines are the CDCC and CRC calculations. The dashed lines and circles are, respectively, the DWBA calculations and the experimental data from Ref.~\cite{Esc07}.}
\end{center}
\end{figure}

The calculated energy and angular distributions of the 
$\alpha$ particles are compared with the data in Figs.~\ref{dsde_crc} and \ref{dsdw_crc}, respectively. For comparison, the DWBA calculations
of Ref.~\cite{Esc07} are also included (dashed lines).  We see that 
the  CRC calculations, represented by solid lines, reproduce the data with a similar quality of the DWBA calculations, 
although the magnitude is somewhat underestimated. This discrepancy might be due to the limitations of our di-neutron model (for both the projectile and target states), to the choice of the 
underlying interactions, or to the contribution of other channels. Considering the simplicity of our model, we can say that the overall agreement is 
fairly good.    Therefore, unlike DWBA calculations, the CRC calculations are able of reproducing simultaneously the elastic and $\alpha$ production data without requiring any phenomenological fit of the elastic cross section.

\begin{figure}[tb]
\begin{center}
 {\centering \resizebox*{0.9\columnwidth}{!}{\includegraphics{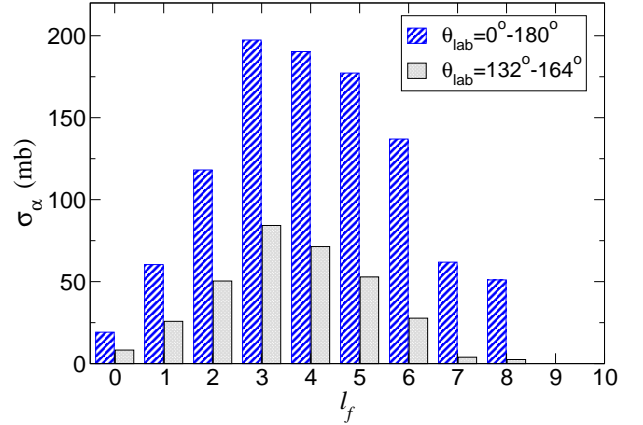}} \par}
\caption{\label{bu_vs_lf} Two-neutron transfer cross section as a function of the 2n-\nuc{208}{Pb} orbital angular momentum $l_f$, for the CRC calculations performed at an incident energy of 22 MeV. The dashed 
bars  and  solid bars correspond, respectively, to the full angular range and the integrated angular distribution for $\theta_\mathrm{lab}=132^\circ-164^\circ$.}
\end{center}
\end{figure}

For a meaningful comparison with the data, we have verified that the number of partial waves used for the 2n-\nuc{208}{Pb} motion ($l_f=0-8$) provides also convergence of the $\alpha$ cross section in the angular range where data exist. This is 
shown in Fig.~\ref{bu_vs_lf},  where we plot the contribution of each $l_f$ to the angle-integrated transfer cross section for $E_\mathrm{lab}$=22~MeV. 
The dark histograms correspond to the angular range 
$\theta_\mathrm{lab}=132^\circ-164^\circ$, whereas the dashed bars are for the full angular range. It is clear from this figure that the model space used in the CRC calculations yields convergence within the angular range of interest and hence 
we conclude that this model space  is suitable to describe the  $\alpha$ particles at 
backward angles. At forward angles, the histograms of Fig.~\ref{bu_vs_lf} suggest a non-negligible contributions from $l_f > 8$ waves.  Although there are no data for the $\alpha$ particles emitted at these forward angles, it is expected that these 
are more easily described by the CDCC calculations.  

\section{Trivially-equivalent local polarization potentials}
The analysis of the preceding section demonstrates that the two-neutron stripping channel can explain  the long-range effect found 
in previous phenomenological analyses of the elastic data  \cite{Kak03,Kak06,San08}. To corroborate this conclusion in a more quantitative way, we have 
evaluated the so called {\it trivially-equivalent local polarization potential} (TELP). This is a local and $L$-independent 
potential constructed from the solution of the coupled equations in such a way that it simulates the net effect of the couplings on the 
elastic scattering. Here, the TELP is calculated following the prescription proposed by Thompson {\it et al.}~\cite{Tho89}, 
which involves two steps. First, for each total angular momentum, a local polarization potential is calculated from the source term
of the elastic channel equation. Then, a $L$-independent potential is constructed by averaging these  $L-$dependent polarization potentials, 
weighted with the transfer/breakup  cross section for  each angular momentum. 
The TELP obtained by this procedure should be regarded as a  simplified local approximation of a  complicated coupled-channels system.  If the TELP extracted from the coupled-channels 
calculation is a good representation of the overall effect of the couplings, the solution of the single-channel 
Schr\"odinger equation  
with the effective potential $U_\mathrm{eff}=U_\mathrm{bare}+U_\mathrm{TELP}$  should reproduce the elastic scattering similar to the one obtained with the full coupled-channels calculation. 

\begin{figure}[tb]
\begin{center}
 {\centering \resizebox*{0.9\columnwidth}{!}{\includegraphics{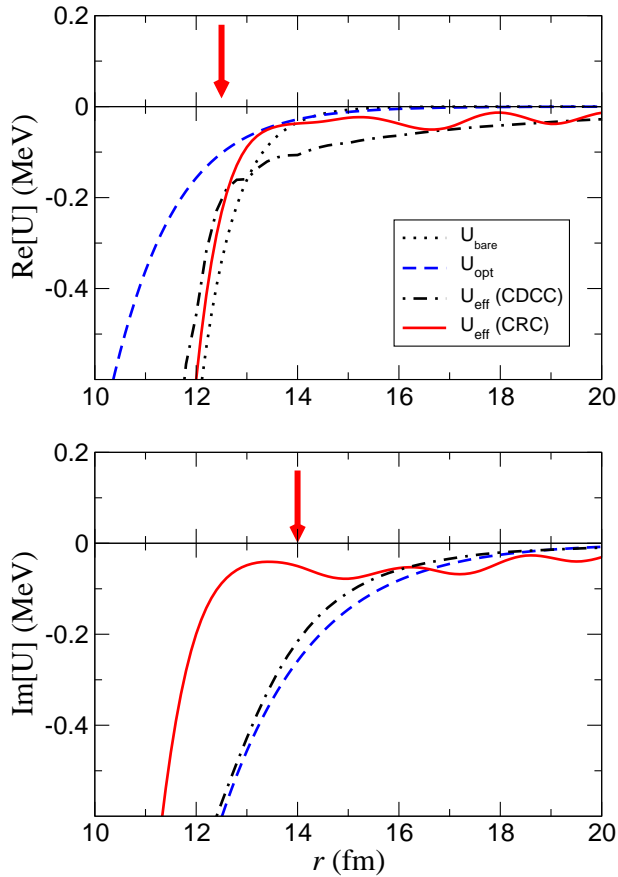}} \par}
\caption{\label{veff} Effective potential,  $U_\mathrm{eff=}U_\mathrm{bare}+U_\mathrm{TELP}$,  for \hepb at 22 MeV, extracted from 
the CDCC (dot-dashed lines) and CRC calculations (solid lines). The dashed line is the phenomenological optical model potential 
extracted in Ref.~\cite{San08} from the fit of the elastic data. The dotted line is the bare interaction used in the CRC 
calculations. The top and bottom panels correspond to the real and imaginary parts, respectively. The inset in the top figure compares  the  polarization potential $U_\mathrm{TELP}$ extracted from the CDCC and CRC calculations with the bare interaction.}
\end{center}
\end{figure}

In Fig.~\ref{veff} we show the effective potentials ($U_\mathrm{eff}$) extracted from the CDCC  and CRC  calculations at $E_\mathrm{lab}=22$ MeV. The top and bottom panels correspond to the real and imaginary parts, respectively. The arrows indicate the radius of sensitivity of the 
real and imaginary parts of the optical potential, according to the OM analysis performed 
in \cite{San08}. 


The dot-dashed and solid lines correspond, respectively, to the effective potential $U_\mathrm{eff}$ extracted from the 
CDCC and CRC  calculations. For comparison, we have included also the double-folding SPP potential (dotted line in top panel), which corresponds to the bare interaction in the CRC calculations, and the phenomenological optical potential 
extracted in Ref.~\cite{San08} from the fit of the elastic data (dashed line). 


For distances $r\gtrsim 15$ fm, the real part of the effective potential (in both the CRC and CDCC calculations) is dominated by a long-range attractive tail. This  attractive component is known to arise from the dipole Coulomb interaction
 \cite{Rus09,Mac09,jp10}. Although the direct breakup model provides a more natural representation for this effect, it is noticeable that the TELP extracted from the CRC shows up also this behaviour.

At distances around the strong absorption radius the TELP becomes repulsive, making the effective potential   shallower  than the bare interaction. This repulsive component is mainly due to nuclear couplings \cite{jp10}. The OM potential is also less intense than the bare interaction, but it does not exhibit the long-range tail observed in the TELPs. This might be due to the fact that the real part of the optical potential is mostly sensitive to distances around the strong absorption radius. 

In the bottom panel of Fig.~\ref{veff} we see 
that the imaginary part of the effective potential extracted from both, the CDCC and CRC calculations,  is absorptive and exhibits also a diffuse tail. This 
behaviour is also observed in the optical potential (dashed line) and is related to the long-range absorption effect 
discussed in previous optical model analyses of this reaction \cite{Kak03,Kak06,San08}.  Therefore, another important conclusion of this work is that the CDCC and CRC calculations provide a  microscopic interpretation of this long-range absorption effect in terms of channel couplings.   Quantitatively, the effective potential extracted from the CDCC calculations is closer to the phenomenological optical potential. We emphasize, however, that due to the $L$ average, the TELP is not strictly 
equivalent to the full set of equations, and hence only the qualitative behavior is meaningful. In addition, the transfer couplings are 
intrinsically non-local and thus their approximation by a local object has to be interpreted with  caution. 

\section{Summary and conclusions}
We have studied the elastic and $\alpha$-production channel for the \nuc{6}{He}+\nuc{208}{Pb} reaction at energies around the Coulomb 
barrier. The available experimental data have been compared with three-body Continuum-Discretized Coupled-Channels (CDCC) and 
Coupled-Reaction-Channels (CRC) calculations using in both cases a di-neutron model of the $^6$He nucleus.  

In the CDCC calculations, the projectile dissociation is taken into account by including the coupling with the \nuc{4}{He}+2n continuum states.  The inclusion of these couplings have a significant effect on the calculated elastic cross section,  and provides a good description of the experimental elastic data. However, the calculated $\alpha$ cross sections largely underestimate the experimental data, in agreement with the findings of Ref.~\cite{Esc07}. 

In the CRC calculations, we consider the coupling to 2n transfer channels and its influence on the elastic scattering cross section. For the bare interaction, we have used 
the microscopic double-folding S\~ao Paulo potential, supplemented with a short-range Woods-Saxon imaginary potential to account 
for complete fusion. The inclusion of two-neutron transfer channels produces a strong effect on the elastic cross section, providing a very good agreement with the experimental data. Moreover, these calculations explain also reasonably well the energy and angular distribution of the $\alpha$ particles measured at backward angles. In particular, the shape of the energy distribution is very well accounted for, although the 
magnitude of the cross section is somewhat underestimated.  This discrepancy might be due to the limitations of our simple di-neutron 
model or the choice of the interactions. In addition,
besides the two-neutron transfer channels, other channels could contribute to this reaction. 
For example, in Ref.~\cite{Kee08} the authors studied the effect of the one-neutron stripping channel, (\nuc{6}{He},\nuc{5}{He}), on the elastic and 
fusion  cross section for several targets. In the \hepb case, they found that the inclusion of these channels produce a reduction of 
the elastic cross section in the region of the Coulomb rainbow, and an increase at backward angles. The simultaneous inclusion of 
both mechanisms would be of interest.

The trivially-equivalent 
local polarization potentials (TELP) derived from the  CRC and CDCC calculations exhibit the characteristic long-range real and 
absorptive parts, in agreement with the behavior observed in the phenomenological optical potentials extracted in 
previous OM analyses of the same data. 


%
%
\section*{Acknowledgements}
This work has been supported by the Spanish Ministerio de Ciencia e Innovaci\'on
under projects FPA2009-08848, FPA2009-07653, by the local government of Junta de Andaluc\'{\i}a under project P07-FQM-02894 and the Spanish Consolider-Ingenio 2010 Programme CPAN (CSD2007-00042). 




\section*{References}
\bibliographystyle{elsarticle-num}
\bibliography{he6pb_crc}







\end{document}